\documentclass{ws-ijmpe}
\usepackage[super,compress]{cite}

\newcommand{\zaa}{{\em Astron.~Astrophys.}}

\newcommand{\zapj}{{\em Astrophys.~J.}}

\newcommand{\zapjs}{{\em Astrophys.~J.~S.}}

\newcommand{\znat}{{\em Nature}}

\newcommand{\znpa}{{\em Nucl.~Phys. A}}

\newcommand{\zplb}{\emph{Phys.~Lett. B}}
\newcommand{\zpr}{{\em Phys.~Rev.}}
\newcommand{\zprep}{{\em Phys.~Rep.}}
\newcommand{\zprd}{{\em Phys.~Rev. D}}
\newcommand{\zprc}{{\em Phys.~Rev. C}}
\newcommand{\zprl}{{\em Phys.~Rev.~Lett.}}

\newcommand{\zadndt}{{\em At. Data Nucl. Data Tables}}
\newcommand{\zmnras}{{\em Mon. Not. R. Astron. Soc.}}

\newcommand{\zrmp}{{\em Rev. Mod. Phys.}}
\newcommand{\zjcap}{{\em J. Cosmology Astropart. Phys.}}
\newcommand{\zetal}{\emph{et al.}}
\newcommand{\zjpg}{{\it J. Phys. G}} 

\newcommand{\obh}{$\Omega_{\mathrm{b}}{\cdot}h^2$}

\newcommand{\deu}{${\rm D}$}
\newcommand{\tro}{$^3{\rm He}$}
\newcommand{\qua}{$^4{\rm He}$}
\newcommand{\six}{$^{6}{\rm Li}$}
\newcommand{\sep}{$^{7}{\rm Li}$}

\newcommand{\neu}{$^{9}$Be}
\newcommand{\dix}{$^{10}$B}
\newcommand{\onz}{$^{11}$B}

\newcommand{\hli}{$^4$He, D, $^3$He and $^{7}$Li}

\newcommand{\ddn}{D(d,n)$^3$He}
\newcommand{\ddp}{D(d,p)$^3$H}
\newcommand{\dpg}{D(p,$\gamma)^3$He}
\newcommand{\hag}{$^3$He($\alpha,\gamma)^7$Be}

\newcommand{\bedp}{$^7$Be(d,p)2$\alpha$}
\newcommand{\bery}{$^{7}$Be}

\newcommand{\etal}{et al.}

\newcommand{\bbn}{big bang  nucleosynthesis}
\newcommand{\sbbn}{standard big bang nucleosynthesis}

\begin{document}

\markboth{A. Coc and E. Vangioni}{Primordial Nucleosynthesis}

\catchline{}{}{}{}{}

\title{Primordial Nucleosynthesis}

\author{Alain Coc}

\address{{Centre de Sciences Nucl\'eaires et de Sciences de la
Mati\`ere (CSNSM), CNRS/IN2P3, Univ.~Paris-Sud, Universit\'e Paris--Saclay, 
B\^atiment 104, F--91405 Orsay Campus, France}\\
coc@csnsm.in2p3.fr}

\author{Elisabeth Vangioni}
\address{Institut d'Astrophysique de Paris,
              UMR-7095 du CNRS, Universit\'e Pierre et Marie
              Curie,\\
              98 bis bd Arago, 75014 Paris (France),\\
              Sorbonne Universit\'es, Institut Lagrange de Paris, 98 bis bd Arago, 75014 Paris (France)\\
              vangioni@iap.fr}

\maketitle

\begin{history}
\received{\today}
\revised{Day Month Year}
\end{history}

\begin{abstract}
Primordial nucleosynthesis, or \bbn\ (BBN), is one of the three evidences 
for the big bang model, together with the expansion of the universe
and the Cosmic Microwave Background. There is a good global agreement over a 
range of nine orders of magnitude between
abundances of \hli\  deduced from observations, and calculated in primordial
nucleosynthesis. 
However, there remains a yet--unexplained discrepancy of a factor $\approx$3, between  the calculated 
and observed lithium primordial abundances,
that has not been reduced, neither by recent nuclear physics experiments, nor by new observations. 
The precision in deuterium observations in cosmological clouds has recently improved dramatically, so 
that nuclear cross sections involved in deuterium BBN need to be known with similar precision.
We will shortly discuss nuclear aspects related to BBN of Li and D, BBN with non-standard neutron sources,
and finally, improved sensitivity studies using Monte Carlo that can be used in other sites of nucleosynthesis.
\end{abstract}

\keywords{Big Bang Nucleosynthesis; Nuclear reactions; Cosmology.}

\ccode{PACS numbers: 26.35.+c, 98.80.Es, 98.80.Ft}


\section{Introduction}
\label{s:intro}

Besides the universal expansion and the cosmic microwave background (CMB) radiation, the third evidence for 
the hot big--bang model comes from primordial or big bang nucleosynthesis (BBN).
During the first $\approx$20 minutes of the universe, when it was dense and hot enough for
nuclear reactions to take place, BBN produced the so called  ``light elements'',  \hli, together with   
only minute traces of heavier nuclei. 
There is indeed a good overall agreement between primordial 
abundances of the light elements either deduced from observations or from primordial
nucleosynthesis calculations. 

It is worth reminding that {\em standard} BBN relies on textbook physics and that,  
with the exception of the baryonic density, all BBN parameters have now been
determined by laboratory measurements. 
However, in the past, BBN has been essential, to first estimate the baryonic density 
of the universe, $\rho_{\rm B} = (1-3)\times10^{-31}$~g/cm$^3$ \cite{Wag73}, and to give an upper limit 
on the  number of neutrino families $N_\nu\leq3$\cite{Yan79}.
The number of light neutrino families is now known from the measurement of the $Z^0$ 
width by LEP experiments at CERN: $N_\nu$ = 2.9840$\pm$0.0082~\cite{LEP}. 
The observations of the anisotropies of CMB by space missions, in particular WMAP \cite{WMAP9} and {\it Planck} \cite{Planck13,Planck15}, 
have enabled the extraction of cosmological parameters.
It  includes the baryonic density of the universe which is now
measured at better than the percent level, a precision that cannot be matched by BBN.  
The other quantities that enter into BBN calculations are in the nuclear physics sector and strongly
constrained by laboratory experiments.
The nuclear reaction rates affecting the production of the $A<8$ isotopes have all been measured in nuclear physics 
laboratories or can be calculated.
Hence, there is no more free parameter in standard BBN  and the calculated 
primordial abundances are in principle only affected by the moderate uncertainties in
nuclear cross sections. 
When calculated primordial abundances are compared with astronomical observations in primitive astrophysical sites,
agreement is generally good. However,  there is a discrepancy for \sep\ that has not yet 
found a consensual explanation. Besides, the recent improved precision on deuterium observations demands equivalent 
progress in some nuclear cross sections.

Accordingly, present \bbn\ studies are focused on $i$) solving the {\em lithium problem}, $ii$) improving
the accuracy of the predictions to match increasing precision on observations and $iii$) probe the physics
of the early universe. 
Indeed, when we look back in time, it is the ultimate process for which, {\it a priori}, we know all the 
physics involved:  departure from its predictions could provide hints 
or constraints on new physics or astrophysics \cite{Ioc09,Pos10}.

\section{Thermal history of the universe}
\label{s:history}

In order to perform the nucleosynthesis calculation one first needs to know the time evolutions of the baryonic density
and temperature. They are obtained from the rate of expansion of the universe and thermodynamic considerations.  
Assuming homogeneity and isotropy, the geometry of the universe is described by the
Friedmann--Lema\^{\i}tre--Robertson--Walker metrics:
\begin{equation}
ds^2=dt^2-a^2(t)\left(\frac{dr^2}{1-kr^2}+
r^2(d\theta^2+\sin{\theta}d\phi^2)\right),
\label{q:metric}
\end{equation} 
where $a(t)$ is the scale factor, describing the expansion, and $k$ = 0 or $\pm$1 marks the absence or sign of space curvature.   
Using the Einstein equation that links the curvature and energy--momentum tensors leads to the 
Friedmann equation that links the rate of expansion 
[$H(t)$]
to the energy density:
\begin{equation}
H^2(t)\equiv\left(\frac{\dot{a}}{a}\right)^2=
\frac{8{\pi}G(\rho_\mathrm{R}+\rho_\mathrm{M})}{3}-\frac{k}{a^2}+\frac{\Lambda}{3},
\label{q:fried}
\end{equation} 
where $G$ is the gravitational constant, $\rho_\mathrm{M}$ is the non-relativistic matter density, $\rho_\mathrm{R}$ the radiation density and 
$\Lambda$ the cosmological constant (see e.g. Weinberg \cite{Weinberg2008}).

When considering the density components of the universe, it is convenient
to refer to the {\em critical density} which corresponds to a flat (i.e. Euclidean) space. 
It is given by [$k=0$, $\Lambda=0$ in Eq.~(\ref{q:fried})]:
\begin{equation}
\rho_{0,C}={{3H_0^2}\over{8{\pi}G}}=1.88\;h^2\times10^{-29}\;
\mathrm{g/cm^3}\;\mathrm{or}\;2.9\;h^2\times10^{11}\;
\mathrm{M_\odot}/\mathrm{Mpc^3},
\label{q:critic}
\end{equation}
where $H_0=h\times$100~km/s/Mpc , is the Hubble constant with $h$ $\approx$ 0.68 \cite{Planck15}.
It corresponds to a density of
a few hydrogen atoms per cubic meter or one typical galaxy per cubic
megaparsec. Densities are usually given relatively to $\rho_{0,C}$ with the 
notation $\Omega\equiv\rho/\rho_{0,C}$.
The total density is very close to the critical density and is dominated by vacuum energy and dark matter contributions while the baryonic matter only 
amounts to $\approx$5\% of the total density, or 16\% of the total matter 
content. 
What we can observe with our telescopes, because it emits light,
corresponds to only $\sim10^{-3}$ of the total density \cite{Fuk04}. 
CMB observations lead to 
 \cite{Planck15}\footnote{We consider the constraints obtained with the largest set of data (TT,TE,EE+lowP) without any external data.
}: 
\begin{equation}
\Omega_b{\cdot}h^2=0.02225\pm0.00016.
\label{q:omegab}
\end{equation}

Nevertheless, in spite of its modest contribution, baryonic matter is important as this is the only one we know and observe. 
The corresponding baryonic density, calculated from Eqs.~(\ref{q:omegab}) and (\ref{q:critic}) is $\rho_{\rm B}\approx4\times10^{-31}$~g/cm$^3$,
only slightly above the first evaluation \cite{Wag73}.
It is \obh\ that is used directly in BBN calculations and is provided by CMB analyses, but it is usual to introduce $\eta$,  the baryon--to--photon number ratio, 
which remains constant during the expansion (after electron--positron annihilation), and is directly related to the baryonic density relative to 
the critical density by $\eta$ = 2.7377$\times10^{-8}$  \obh\ \cite{Coc14b}.

\begin{figure}[htb]
\begin{center}
\includegraphics[width=.8\textwidth]{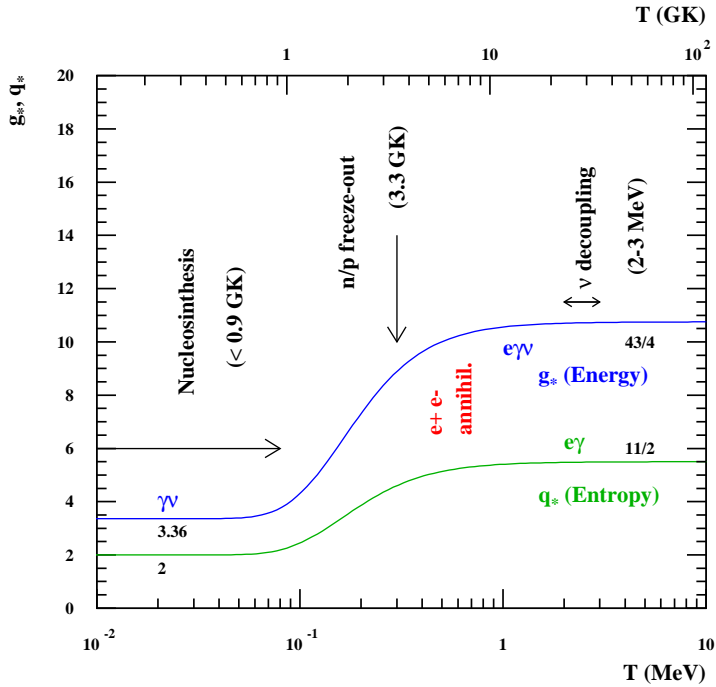}
\caption{The spin factors $g_*$ and $q_*$ appearing in Eqs.~(\ref{q:expan}) and (\ref{q:entro}), as a function of temperature, 
together with the landmarks of BBN.} 
\label{f:geff}
\end{center}
\end{figure}

However, at the BBN epoch, the main contributions to the energy density, that govern the expansion rate,  are quite different from the present ones.  
During the expansion, the non-relativistic (dark and baryonic) matter component of the density is diluted according to  
$\rho_\mathrm{M}\propto a^{-3}$, while for relativistic particles ("radiation"), there is an additional factor due to redshift
and  $\rho_\mathrm{R}\propto a^{-4}$. The two other terms in the right hand side of Eq.~\ref{q:fried} scale as  $a^{-2}$ (curvature)
and  $a^{0}$ (cosmological constant $\Lambda$).
The important consequence is that during BBN,  when $a$ is $\approx10^8$ times smaller than today, $H(t)$ is only governed
by relativistic particles while the baryons, cold dark matter, 
cosmological constant or curvature terms play no 
role.
Eq.~(\ref{q:fried}) becomes:
\begin{equation}
\frac{1}{a}
\frac{\mathrm{d}a}{\mathrm{d}t}=
\sqrt{{{8{\pi}G}\over3}\mathrm{a_R}\frac{g_*(T)}{2}}{\times}T^2,
\label{q:expan}
\end{equation} 
where we have used  the Stefan-Boltzmann law $\mathrm{a_R}T^4$ for the radiation energy density. 
At temperatures slightly above 10$^{10}$~K, the present particles are: photons, electrons, positrons, the three families of 
neutrinos and antineutrinos plus a few neutrons and protons. 
The effective spin factor, $g_*$, decreases whenever the temperature drops 
below a mass threshold for the particle--antiparticle annihilation of    
each species. During BBN, only e$^+$ and e$^-$ annihilation has to be considered.
Hence, the contributions to $g_*(T)$ (Fig.~\ref{f:geff}) come from photons, neutrinos 
and electrons/positrons before they annihilate. 
The released energy is shared among the other particles they were in equilibrium with: photons and baryons but not
neutrinos as it happens after their decoupling (Fig.~\ref{f:geff}). 
During the adiabatic expansion the entropy densities of neutrinos [Eq.~(\ref{q:entronu})] and photons+electrons [Eq.~(\ref{q:entro})] stay 
separately constant: 
\begin{equation}
a^3T_\nu^3=\mathrm{Cste.}
\label{q:entronu}
\end{equation} 
\begin{equation}
a^3q_*^{e\gamma}(T)T_\gamma^3=\mathrm{Cste.}
\label{q:entro}
\end{equation} 
with the temperature dependent (due to e+e- annihilation) spin factor, $q_*$  shown in Fig.~\ref{f:geff}.
Note that the baryonic density does not have at 
this epoch any influence on the rate of expansion of the 
universe (i.e. Hubble parameter). 
Its influence on nucleosynthesis is simply that a higher 
density of nuclei induces a larger number of reactions taking place per unit time.
Apart from thermonuclear reaction rates, the quantities needed for BBN calculations are the photon/ion temperature $T(t)$, the
neutrino temperature $T_\nu(t)$ and the baryonic density $\rho_\mathrm{B}(t)\;{\propto}\;\Omega_b{\cdot}h^2\;a^{-3}(t)$ as a function of time.
They   are simply obtained by numerically solving Eqs.~(\ref{q:expan})--(\ref{q:entro}) \cite{Weinberg2008}. 
Hence, compared to stellar nucleosynthesis, \sbbn\ seems at first as a simple problem without convection, diffusion, or other
mixing mechanism, based on known physics, and with most important reactions cross sections 
measured at the relevant energies.
Other differences are that the density during BBN is orders of magnitude lower than in stellar cores, so that three--body reactions
(the triple--alpha reaction) are hindered which together with unusual abundances of n, d, t and $^3$He lead to different nuclear flows. 

\section{Observed abundances}
\label{s:obs}

During the evolution of galaxies, nucleosynthesis takes place mainly
in massive stars which release matter enriched in heavy elements into the 
interstellar medium when they explode as supernovae. 
Accordingly, the abundance of heavy elements, in star forming gas,
increases with time. The observed abundance of 
{\em metals} (elements heavier than helium) is hence an indication of age: the older the lower the {\em metallicity}.
Primordial abundances are hence extracted from observations of objects 
with very small metallicity.

After BBN, concerning the light cosmological elements, \sep\ can be both produced (spallation, AGB stars, novae
\footnote{Recent observations \cite{Taj15,Izz15} have confirmed Li production by novae, 
at  a level even higher than model predictions \cite{HJCI96}.})
and destroyed (in the {\em interior} of stars). 
The life expectancy of stars with masses lower than our Sun is larger than the age of the 
universe so that very old such stars can still be observed in the halo of our Galaxy.
In this context, lithium can be observed at the surface of these stars and its abundance was found to be
independent of metallicity, below $\approx0.1$ of the solar metallicity. 
This {\em plateau} was discovered by Fran\c{c}ois and Monique Spite \cite{Spi82} and this
constant Li abundance was interpreted as corresponding to the BBN \sep\ production.
The thinness of ``Spite plateau" is an indication that surface Li depletion may not have been 
very effective so that it should reflect the primordial value \footnote{Note that recent lithium 
observations \cite{How12} have been done in the Small Magellanic Cloud which has a quarter of the sun's 
metallicity and a Li abundance nearly equal to the BBN predictions.}.
The analysis of Sbordone et al.  \cite{Sbo10}  gives Li/H=(1.58$\pm0.3)\times10^{-10}$ 
(i.e. number of atoms relative to hydrogen).

\begin{figure}[h]
\begin{center}
\includegraphics[width=.8\textwidth]{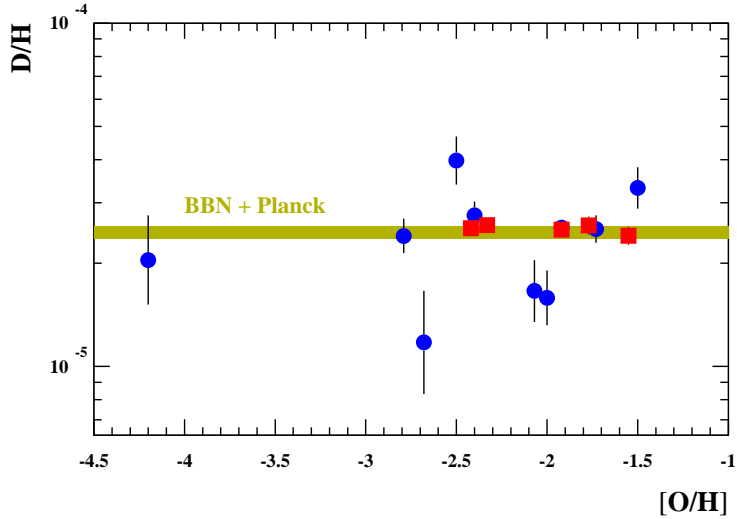}
\caption{D/H observations, as a function of metallicity, from Pettini et al. \cite{Pet12} (blue circles) and Cooke et al. \cite{Coo14} (red squares). These
most recent observations \cite{Coo14} have very small error bars and show very few dispersion, and are in fair agreement with BBN calculations \cite{Coc1X}.}
\label{f:deuobs}
\end{center}
\end{figure}

Deuterium is a very fragile
isotope. It can only be destroyed after BBN throughout stellar evolution (\S~\ref{s:deut}). 
The deuterium abundance closest to primordial abundance is determined from the observation of a few
cosmological clouds at high redshift (Fig.~\ref{f:deuobs}), on the line of sight of distant quasars. 
Recently, Cooke et al.~\cite{Coo14} have made new observations and  reanalyzed existing data, 
that lead to  a new average value of D/H = $(2.53 \pm 0.04) \times 10^{-5}$,
lower and with smaller uncertainties than in previous determinations.  
If such a precision of 1.6\% in observations is confirmed, great care should be paid to 
nuclear cross sections affecting deuterium nucleosynthesis.

After BBN, \qua\ is also produced by stars. Its primordial abundance is 
deduced from observations in H{\sc ii}  (ionized hydrogen) regions of compact 
blue galaxies. 
Galaxies are thought to be formed by the agglomeration of such dwarf galaxies, in a hierarchical structure formation paradigm, 
which are hence considered as more primitive.
To account for stellar production, \qua\  abundance deduced from observations 
is extrapolated to zero, followed by atomic physics corrections. Aver et al.  \cite{Ave15}
obtained   $Y_p=0.2449\pm0.0040$ (in mass fraction).

Contrary to \qua, \tro\ is both produced and destroyed in stars so that
the evolution of its abundance as a function of time is not well known.
Because of the difficulties of helium observations and the small 
$^3$He/$^4$He ratio, \tro\ has only been observed in our Galaxy: 
\tro/H=$ (0.9-1.3)  \times10^{-5}$ \cite{Ban02}.

\section{Nuclear physics aspects}
\label{s:nucl}

At high temperature, neutrons and protons  are in thermal equilibrium so that,  their number ratio is 
$N_n/N_p=\exp(-Q_{np}/k_BT)$ where $Q_{np}$ = 1.29~MeV is the neutron-proton mass difference. 
This holds until  $T\approx$3.3~GK, when the weak rates that govern the n$\leftrightarrow$p reactions 
($\nu_e$+n$\leftrightarrow$e$^-$+p and $\bar{\nu}_e$+p$\leftrightarrow$e$^+$+n),
become slower than the rate of expansion $H(t)$.  
Afterward, the ratio at {\em freezeout} $N_n/N_p\approx$0.17 further decreases to $N_n/N_p\approx$0.13 
due to free neutron beta decay until the temperature is low enough ($T\approx$0.9~GK) for the first nuclear reaction 
n+p$\rightarrow$D+$\gamma$ to become faster than the reverse photodisintegration (D+$\gamma$$\rightarrow$n+p) that  
prevented the production of heavier nuclei. From that point on, the remaining neutrons almost entirely end up bound in \qua\ 
while only traces of \deu, \tro\ and \sep\ being produced. 
Hence, the \qua\ yield is directly related to the $N_n/N_p$ ratio at freezeout that is at the expansion rate $H(t)$
comparable to the weak rates. 

\begin{figure}[htb!]
\begin{center}
\includegraphics[width=.8\textwidth]{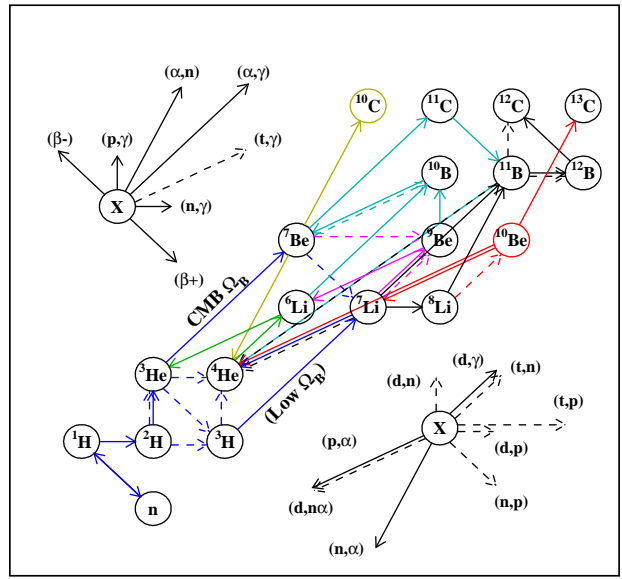}
\caption{Nuclear network  of the most important reactions in BBN, up to\sep\ (blue), 
including \six\ (green), $^{10,11}$B (light blue), \neu\ (pink) and up to CNO (black and red).
The red arrows represents the newly found reactions that could affect CNO production.
The yellow arrows indicate the \bedp\ and \bery+\tro\ reactions that were considered as possible
extra \bery\ destruction mechanisms.} 
\label{f:netw}
\end{center}
\end{figure}

Figure~\ref{f:netw} displays the most important reactions for the BBN up to CNO, however, only a dozen 
are important for \hli\ production.     
There are many other reactions connecting these isotopes, but their cross sections are too small and/or 
the reactants too scarce to have any significant effect. 

The weak reaction rates, involved in n$\leftrightarrow$p equilibrium,  come from the standard theory
of the weak interaction. They are calculated \cite{Dic82} with, as only experimental input, the neutron lifetime 
whose experimental value, 880.3$\pm$1.1~s,  \cite{PDG} is still a matter of debate \cite{Wie11,You14} that affects 
directly the \qua\ production \cite{Mat05}. 
The n+p$\rightarrow$D+$\gamma$ reaction rate \cite{And06} is also obtained from
theory but in the framework of Effective Field Theory, in good agreement with experiments. 
For the ten remaining reactions, $^2$H(p,$\gamma)^3$He, ~~$^2$H(d,n)$^3$He, 
~$^2$H(d,p)$^3$H, ~$^3$H(d,n)$^4$He, 
~$^3$H($\alpha,\gamma)^7$Li, ~$^3$He(d,p)$^4$He, ~$^3$He(n,p)$^3$H, 
~$^3$He($\alpha,\gamma)^7$Be, ~$^7$Li(p,$\alpha)^4$He and ~$^7$Be(n,p)$^7$Li,
cross sections have been measured in the laboratory at the relevant energies (a few 100~keV). 
This is possible because of the higher energies, hence cross sections, compared to 
typical stellar nucleosynthesis. 
Recent compilations of experimental nuclear data to determine thermonuclear
rates for  BBN, and associated rate uncertainties,  were performed by Descouvemont et al. \cite{Des04}, 
Cyburt \cite{Cyb04} and Serpico et al. \cite{Ser04} for A$\leq$7
and by Coc et al. \cite{Coc12a} for 7$<$A$\leq$12.

\begin{figure}[h]
\begin{center}
\includegraphics[width=.8\textwidth]{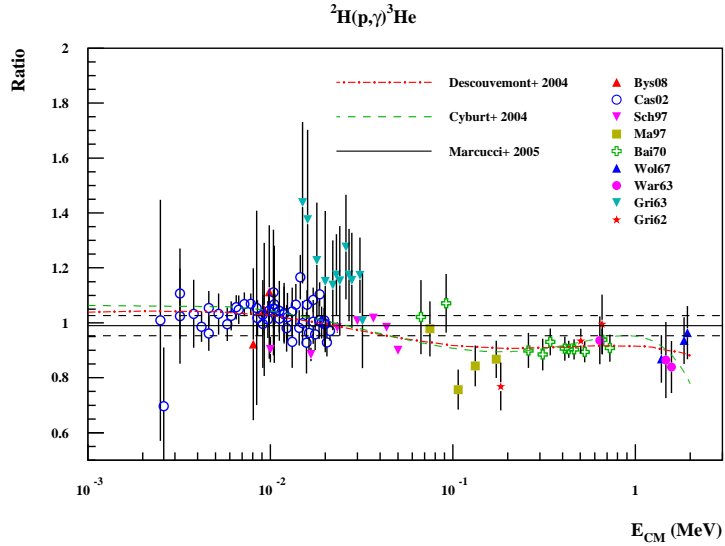}
\caption{Ratio of experimental $S$--factors to the theoretical one \cite{Mar05}, itself normalized to a subset of the experimental data 
\cite{Coc1X}.
Ratio of previous fits \cite{Des04,Cyb04} are driven below theory by the scarce data at BBN energies.}
\label{f:reval}
\end{center}
\end{figure}

Since these A$\leq$7 evaluations, new experimental data (Di Leva et al. \cite{DiL09} and references therein)
has improved the accuracy and reliability of  the important reaction $^3$He($\alpha,\gamma)^7$Be rate \cite{Cyb08a,deB14}.
With the increased precision on deuterium observations, the reactions that govern its nucleosynthesis have also been 
re--investigated \cite{DiV14,Coc1X}. 
Sensitivity studies (e.g. Ref.~\refcite{Cyb04,CV10}) have shown that the $^2$H(d,n)$^3$He, $^2$H(d,p)$^3$H and $^2$H(p,$\gamma)^3$H
reactions, are the most influential on D/H predicted abundance. Since the last dedicated BBN evaluations of reaction 
rates \cite{Cyb04,Ser04,Des04} a new experiment was performed by Leonard \zetal\ \cite{Leo06}.  They measured both the
$^2$H(d,n)$^3$He, $^2$H(d,p)$^3$H cross section between $\approx$50--300~keV, i.e. well within BBN energy range,   
with a quoted uncertainty of 2\%$\pm$1\%. On the contrary, no new experiment concerning the $^2$H(p,$\gamma)^3$H
reaction has been conducted so that its rate uncertainty (5\%--8\% \cite{Des04}), according to  Di Valentino et al. \cite{DiV14},
now dominates the error budget on D/H predictions.  
For instance,  a global increase by an  $\approx1.10\pm0.07$ factor was proposed 
by Di~Valentino et al. \cite{DiV14} and the {\it Planck} collaboration \cite{Planck15} to better match the CMB and D/H observations.
More recently, the \dpg, \ddn\ and \ddp\ experimental data have been used to normalize theoretical $S$--factors \cite{Mar05,Ara11} instead of 
polynomial \cite{Cyb04} or R--Matrix \cite{Des04} fits. For instance Fig.~\ref{f:reval} display available experimental $S$--factors  for the \dpg\ 
reaction divided by the theoretical prediction from Marcucci et al. \cite{Mar05} after normalization (a factor of 0.99) on the data.  
The ratio between the previous fits \cite{Cyb04,Des04} with respect to theory \cite{Mar05} show that they are strongly
affected by the few experimental data available at BBN energies. By using the normalized theoretical $S$--factor instead 
of fits, the \dpg\ rate is found to be higher at BBN temperatures. Together with a similar treatment for the \ddn\ and \ddp\
data, the \deu\ production is naturally reduced (\S~\ref{s:comp}).

Our BBN calculations take advantage of an extended nuclear network \cite{Coc12a} including n, d, t, $^3$He and $\alpha$
induced reactions on targets nuclei up to the CNO region, in order to obtain abundances of \six, \neu, \onz\ and CNO isotopes,
and to take into account sub-leading nuclear flows that may affect \hli\ abundances.
Reaction rates and associated uncertainties, derived from experimental data or from theory (see Ref.~\refcite{Coc12a}),
are used as input for Monte Carlo BBN calculations to evaluate uncertainties on the resulting abundances
and investigate their possible hidden interrelationship with peculiar reactions (\S~\ref{s:stats}).

\section{BBN primordial abundances compared to observations}
\label{s:comp}

\begin{figure}[htb!]
\begin{center}
\includegraphics[height=.8\textheight]{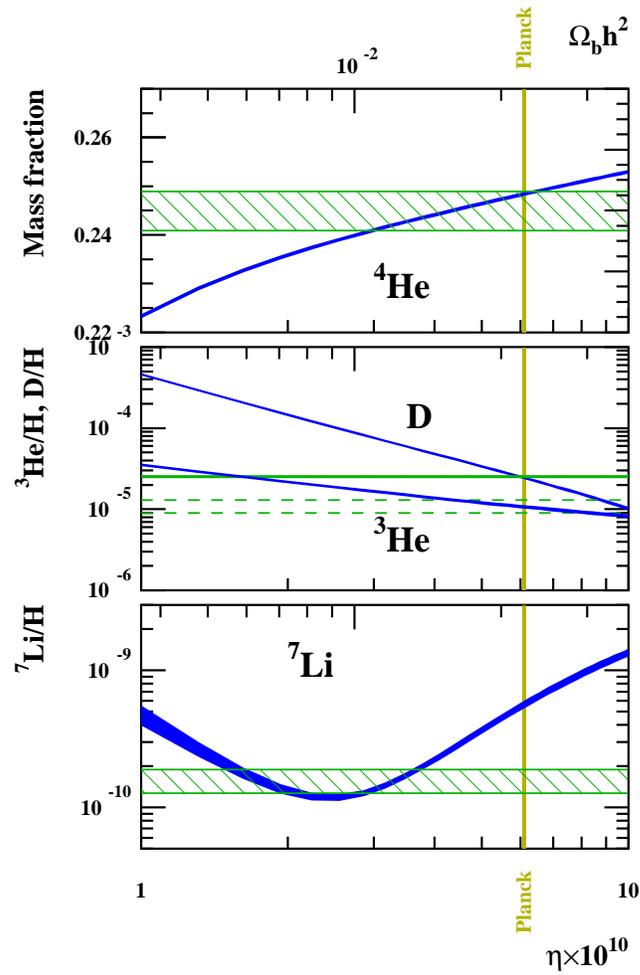}
\caption{Abundances of \qua\ (mass fraction), \deu, \tro\ and \sep\ (by
number relative to H) as a function of the baryon over photon ratio 
$\eta$ or \obh. (Data are from Ref.~\protect\refcite{Coc1X}). 
The vertical stripe corresponds to the CMB baryonic density \cite{Planck15}
while the horizontal hatched area represent the primordial abundances (\S~\ref{s:obs}).}
\label{f:heli}
\end{center}
\end{figure}    
In Figure~\ref{f:heli}  is represented an updated calculation of the abundances of \hli, as a function of the baryonic density
where the thickness of the curves reflect the nuclear rate uncertainties.  
They were obtained  by a Monte--Carlo calculation using the nuclear rate uncertainties from 
Refs.~\refcite{And06,Des04,Coc12a,deB14,Hou15}. 
The horizontal areas represent the 
primordial abundances deduced from observations as discussed above (\S~\ref{s:obs}).
The vertical stripe represents the baryonic density deduced from
CMB observations \cite{Planck15}.
The calculated primordial abundances at Plank baryonic density  are given in Table~\ref{t:heli}. 
We can observe small evolutions with respect to earlier works \cite{Coc14b}; this comes from slight 
evolution of the baryonic density and the neutron lifetime but more effectively from updated reaction rates:
 \hag\ \cite{deB14}, \dpg, \ddn\ and \ddp\ \cite{Coc1X} and $^7$Be(n,$\alpha)^4$He \cite{Hou15}.
\begin{table}[h]
\begin{center}
\caption{Yields at CMB baryonic density.} 
\begin{tabular}{|c|c|c|c|}
\hline
&   Ref.~\refcite{Coc14b} & This work \cite{Coc1X} & Observations\\ 
\hline
$^4$He       &  0.2482$\pm$0.0003 & 0.2484$\pm$0.0002&  0.2449$\pm$0.0040\cite{Ave15}  \\
\deu/H   ($ \times10^{-5})$               & $2.64^{+0.08}_{-0.07}$ & 2.45$\pm$0.05 &  2.53$\pm$0.04 \cite{Coo14} \\
\tro/H    ($ \times10^{-5}$)              & 1.05$\pm$0.03 & 1.07$\pm$0.03 & (0.9--1.3)\cite{Ban02}         \\
\sep/H ($\times10^{-10}$)       &$4.94_{-0.38}^{+0.40}$  & 5.61$\pm$0.26& 1.58$\pm$0.31 \cite{Sbo10}    \\
\hline
\end{tabular}
\label{t:heli}
\end{center}
\end{table}
A noticeable change is observed on D/H which is significantly reduced due to the re--evaluation of the three above mentioned reaction rates. 
As shown on Figure~\ref{f:heli} and Table~\ref{t:heli}, the primordial abundances deduced 
either by BBN at CMB deduced baryonic density, or from observations,
are in good agreement except for \sep, whose calculated abundance is significantly
higher \cite{Cyb08b,Coc14b,Cyb15} (a factor of $\approx$3.5) than the primordial abundance deduced from observations.
The origin of this discrepancy between CMB+BBN and spectroscopic observations remains an open question (\S~\ref{s:lithium}).

Besides those four {\em light elements} heavier isotopes are produced in minute amounts \cite{Coc12a}:
\neu/H$\approx1\times10^{-18}$, \dix/H$\approx3\times10^{-21}$,
and \onz/H$\approx3\times10^{-16}$.  
A special mention should be made for \six, for which the possible existence of a plateau, in the  $^6$Li/H=10$^{-12}$ to 
10$^{-11}$ range (well above the BBN yield of 1.3$\times10^{-14}$)\cite{Ham10}, had been suggested \cite{Asp06} but
has not been confirmed by subsequent observations \cite{Lin13}.
The CNO \sbbn\ production is found to be CNO/H  $(0.96^{+1.89}_{-0.47})\times10^{-15}$ (too low 
to have an impact on Population III stellar evolution) \cite{Coc14b}.

\section{The lithium problem}
\label{s:lithium}

There are many tentative solutions to this problem (nuclear, observational,
stellar, cosmological,...) \cite{Fie11,Spi12}, but none has provided
yet a fully satisfactory solution.
The derivation of the lithium abundance in halo stars with
the high precision needed is difficult and requires
a fine knowledge of the physics of stellar 
atmosphere (effective temperature scale, population of different ionization 
states, non Local Thermodynamic Equilibrium effects and 1D/3D model 
atmospheres).
There is no lack of phenomena to modify the surface abundance of 
lithium: nuclear burning, rotational induced mixing, atomic diffusion, 
turbulent mixing, mass loss,.... 
However, the flatness of the plateau over three decades in metallicity 
and the relatively small dispersion of data represent a real challenge 
to stellar modeling.\footnote{A recent work\cite{Fu15} suggested that pre--main sequence depletion, regulated by photo-evaporation could achieve this goal.}
One also notes that between the BBN epoch and the birth of the now observed 
halo stars, $\approx$1~Gyr has passed. Primordial abundances could have 
been altered during this period.

\subsection{No nuclear physics solution}

Before invoking non--standard solutions to this problem nuclear solutions have been investigated.   
At the baryonic density deduced from CMB observations, \sep\ is produced indirectly by $^3$He($\alpha,\gamma)^7$Be,
that will, much later decay to \sep\ while it is destroyed by  $^7$Be(n,p)$^7$Li(p,$\alpha)^4$He.
The $^3$He($\alpha,\gamma)^7$Be cross section has long been a subject of debate because of 
systematic differences that were found according to the experimental technique. 
This is now settled \cite{Cyb08a,DiL09,deB14} and the associated uncertainty ($\sim$5\%) is very small
compared to the discrepancy.
To solve this problem, within conventional nuclear physics, one has to search for 
other reactions that could lead to  $^7$Li+$^7$Be increased destruction.  
The $^7$Be(d,p)2$\alpha$ reaction (Fig.~\ref{f:netw}) was a prime candidate \cite{Coc04} but subsequent experiments and analyses ruled out this 
possibility  (Kirsebom \& Davids \cite{Kir11} and references therein). 
Extending this search, recent works \cite{Cha11} suggested the possibility of overlooked resonances 
in nuclear reactions involving $^7$Be, the most promising candidate was found to be in the 
$^7$Be+$^3$He$\rightarrow$$^{10}$C channel.  
However, in a recent experiment the upper limits for the presence of new levels in $^{10}$C 
(and $^{11}$C) were found to be too low to have an impact on \sep\ production \cite{Ham13}. 
The natural way of  $^7$Be destruction in BBN occurs through the $^7$Be(n,p)$^7$Li(p,$\alpha)^4$He channel which is limited by
the scarcity of neutrons. 
Hence, since a supplementary reaction, overlooked in previous studies, seems now to be excluded by experiments \cite{Ham13}, 
a peculiar attention should be paid to an enhanced neutron abundance.

\subsection{Non standard neutron injection during BBN}
\label{s:neut}

It was recognized\cite{Ren88,Jed04}, that extra neutron injection would increase $^7$Be destruction by $^7$Be(n,p)$^7$Li(p,$\alpha)^4$He, 
but at the expense of a rise in the abundance of D/H.  Given the new tight constraints on D/H observations (\S~\ref{s:obs}), one may question if the neutron injection 
mechanism is still a valid agent for reducing the cosmological abundance of lithium. Extending the BBN network to $\approx$400 reactions 
has not lead to the identification of any overlooked conventional neutron source \cite{Coc12a}, i.e. an extra neutron producing reaction. Hence, one has to
investigate non standard neutron sources that can be:     

\begin{enumerate}

\item  {\em Particle decay}. This class of models assume the existence of a hypothetical particle $X$ that can decay and produce neutron, i.e.  
$X\to{\mathrm n}+.....$.

\item  {\em Particle annihilation}. These models assume $X+X\to{\mathrm n}+.....$ pair annihilation.

\item  {\em Resonant particle annihilation.} A narrow resonance in the annihilation cross section is present at some energy.

\item  {\em $n-n'$ oscillation}. This model \cite{Ber01} assumes that there is a mirror world from which {\em mirror--neutrons} can oscillate
into our world. The microphysics is considered to be identical in the two sectors, but the temperatures and baryonic densities are different 
in the two sectors \cite{Ber01}.

\end{enumerate}

\begin{figure}[htb]
\begin{center}
\includegraphics[width=.75\textwidth]{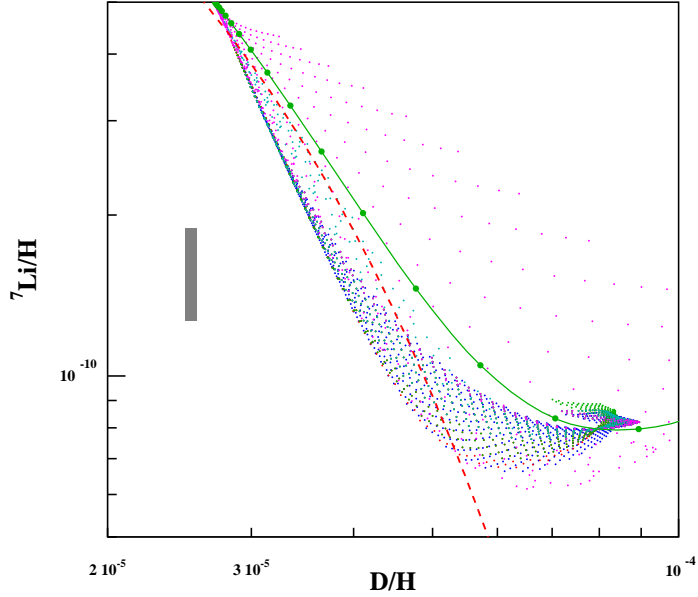}
\caption{Each dot is the prediction of a model \cite{Coc14a} in the space (D/H, $^7$Li/H).
The rectangle corresponds to the D/H observational limits of Ref.~\protect\refcite{Coo14} together with those from Ref.~\protect\refcite{Sbo10} for lithium.
The blue, red and green dots correspond to {\it n-n' oscillation} models the light blue dots correspond to {\it  resonant annihilation} models and the pink dots to {\it particle decay} 
models. The green curve with filled circles  corresponds to the {\it  non--resonant annihilation} model.
The dashed line is a qualitative explanation of this anti--correlation \cite{Coc1X}.
This demonstrates that no model can be in agreement with both lithium-7 and deuterium}
\label{f:neut}
\end{center}
\end{figure}

Figure~\ref{f:neut} is a summary of the results \cite{Coc14a} of BBN calculations within the framework of models 1--4, while varying the relevant parameters. 
Each dot correspond to a set of parameters and different colors correspond to the different models.
It appears that the $^7$Be destruction by the injection of extra neutrons is accompanied by the deuterium over--production, i.e.
that lithium and deuterium abundances are anti--corellated as seen, e.g., in an instantaneous neutron injection model 
(Fig.~4 in Kusakabe et al. \cite{Kus14}) or in a massive gravitino decay model (Fig.~1 in Olive et al. ~\cite{Oli12}).
The reason for this anti--correlation is that besides \bery\ destruction by $^7$Be(n,p)$^7$Li, late time neutron injection, {\em unavoidably} generate
extra deuterium by the $^1$H(n,$\gamma$)D reaction \cite{Kus14,Coc1X}. Neglecting all other reactions, at the relevant temperature
when \bery\ is formed, one obtains \cite{Coc1X} the dashed curve in Fig.~\ref{f:neut}.
The lower limit for lithium abundance appears at Li/H$\approx6\times10^{-11}$ (Fig.~\ref{f:neut}). It comes from the enhanced $^3$H production [by $^3$He(n,p)$^3$H]
that feeds the $^3$H($\alpha,\gamma)^7$Li branch (``Low $\Omega_\mathrm{B}$'' on Fig.~\ref{f:netw}) while \sep\ may not be efficiently destroyed anymore 
by $^7$Li(p,$\alpha)^4$He, because of the lower temperature.

\section{Deuterium cosmic evolution}
\label{s:deut}

Starting from our new BBN D/H value  at redshift of $z\approx10^8$, it is interesting to
follow the cosmic deuterium evolution. This isotope is a good tracer of stellar formation since 
it can only be destroyed from the BBN epoch due to its fragility (burnt at $T > 10^5$  K, 
deuterium is destroyed throughout the cosmic evolution). 

The observational constraints on D/H evolution with redshift, besides the cosmological data from damped Lyman-$\alpha$ (DLA) systems 
at $z\approx2 - 3$, already discussed in Section 3, come from the local D/H observations at present day ($z=0$).
Prodanovic et al. \cite{Pro10} estimate the interstellar medium (ISM) deuterium abundance to be D/H~$ >(2.0\pm 0.1)\times10^{-5}$, 
leading to an astration factor $f_D$ (which is the ratio, D$_\mathrm{BBN}$/D$_\mathrm{present}$) , of $f_D < 1.26\pm 0.1$. 
Linsky  et al. \cite{Lin06} study  reveals a very wide range of observed D/H ratio in the local galactic disk and give as
the most representative value D/H $ > (2.31\pm 0.24)\times10^{-5}$, leading to an astration factor, $f_D$, of less than 1.1.

\begin{figure}[htb]
\begin{center}
 \includegraphics[width=.8\textwidth]{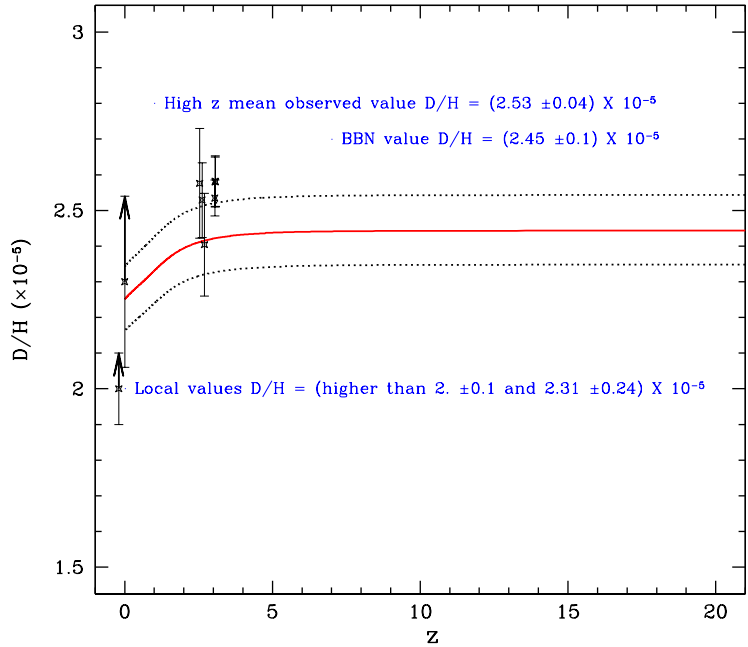} 
\caption{Cosmic D evolution as a function of redshift. 
The red solid curve corresponds to the evolution of D/H using our mean BBN value whereas the black dotted curves correspond to 
the higher and lower 2$\sigma$ limits.  High~$z$ DLAs observations 
come from Cooke et al. \cite{Coo14} whereas local observations come from Linsky et al.  \cite{Lin06} and Prodanovic et al.  \cite{Pro10}} 
\label{f:devol}
\end{center}
\end{figure}

Cosmic chemical evolution depends 
on a stellar initial mass function (IMF) and a star formation rate (SFR). 
Their convolution is measurable through the total observed luminosity density.
Recently, improvements have been made in our understanding
of the global star formation history, particularly at high redshift. 

SFR  evolution with redshift is constrained by many observations.
Specifically, recent data from high redshift galaxy observations (the
Hubble Ultra Deep Field) have significantly extended the
range of redshifts for its determination, from z  = 4 up to
10 \cite{oesch14, bou14}.  It is a key ingredient to all evolution models.
For a comprehensive discussion of these observational advances, see
Bouwens et al.  \cite{bou15}.

In this context we consider the cosmic evolution of D/H
in a cosmological way in the light of the new, somewhat
low, D primordial value derived here.  
 Indeed, we follow its cosmic evolution
 using a model developed in Refs.~\refcite{Dai06,Rol09,Van14}, based
on a hierarchical model for structure formation (Press and Schechter formalism \cite{press74}), 
and we determine the rate at which structures accrete mass. 
The model follows the evolution of the amount of baryons in stars, in structures (ISM) and in the intergalactic medium (IGM).  
The model includes also a description of mass exchanges
between the IGM and ISM (structure formation, galactic outflows), and between the ISM and the stellar
component (star formation, stellar winds and supernova explosions). 

Once the cosmic SFR is specified, several
quantities are obtained as a function of redshift,
namely the abundances of chemical elements, SN rates, reionization of the universe, and more
specifically deuterium. 
Deuterium destruction is governed by low mass stars (since the gas is essentially trapped in these stars and is released only, long after, at low redshift)
whereas metallicity production (elements others than H and He) is governed by high mass stars which reject enriched matter 
at high redshift. A weak destruction of D is consequently not incompatible with a significant formation of heavy elements. 

We consider here the results of the best model described in Ref. \refcite{Van14}, including
a standard mode of Pop II/I stars formation between 0.1 and 100 solar masses. 
The IMF slope is set to the Salpeter value, i.e. $x  = 1.35$ \cite{Sal55}.
 Figure~\ref{f:devol} shows the evolution of D/H as a function
of redshift starting from our 2$\sigma$  BBN limits. Black dotted curves correspond to the present BBN limits, whereas the red solid line corresponds
to the mean. The resulting astration factor is $f_D =1.1$. This cosmic evolution is in overall agreement
with the observed values, implying that the mean abundance of deuterium has only been reduced by a factor of 1.1 to 1.25 (see Ref.~\refcite{Coc1X}) since its formation.
Note however that a tension exists between the BBN D/H value and the
high $z$ measurements leaving clearly little room for a high astration factor.

\section{Statistical methods in (big bang) nucleosynthesis}
\label{s:stats}

Improved techniques have been developed in nucleosynthesis calculations to evaluate uncertainties
on the final results and to identify the main sources of these uncertainties.
It is of special interest for BBN because of the commensurate uncertainties on calculated and observed abundances,
while the lithium problem (\S~\ref{s:lithium}) remains unsolved. 

\begin{figure}[h]
\begin{center}
\resizebox{.9\textwidth}{!}{%
  \includegraphics{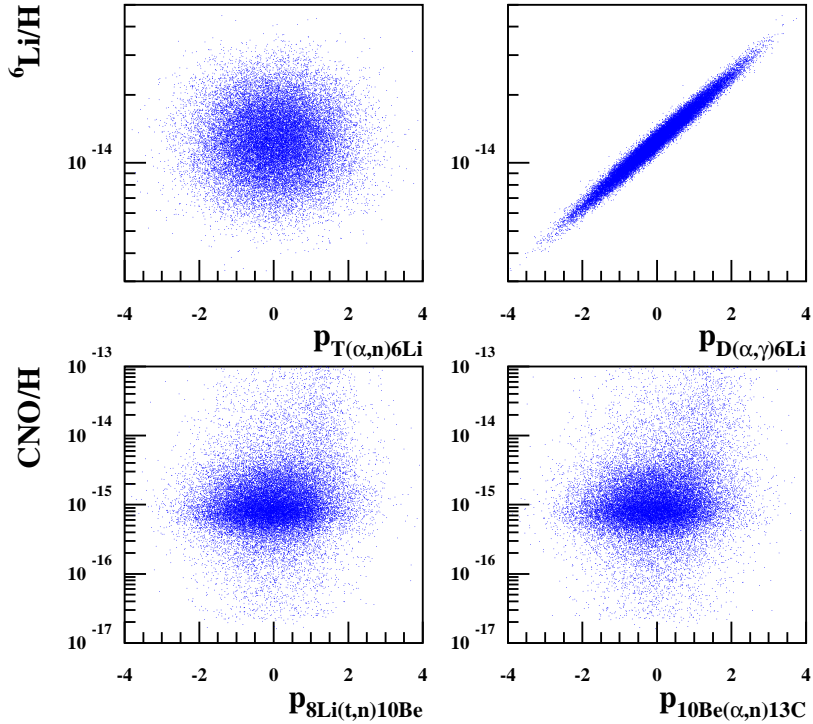}
}
\caption{Top panels: scatter plots of \six\  yields versus random enhancement factors ($p_k$) applied to reaction rates in the context of BBN,
showing no [T($\alpha$,n)$^6$Li] or very strong [D($\alpha,\gamma)^6$Li] correlation with \six.
Bottom panels: scatter plots of CNO/H yields versus random enhancement factors ($p_k$) applied to
reaction rates showing weak correlation  with respectively $^8$Li(t,n)$^{10}$Be  and 
$^{10}$Be($\alpha$,n)$^{13}$C reactions 
(data are from Ref.~\protect\refcite{Coc14b}).}
\label{f:cor}       
\end{center}
\end{figure}

At first, simple sensitivity studies \cite{Cyb04,CV10} (i.e. varying one reaction rate at a time)
have identified the most important BBN reactions for the production of the light elements. 
Unexpected effect can be found this way like the high sensitivity of the \sep\ yield to the 
$^1$H(n,$\gamma)^2$H rate or that an increase of the $^7$Li(d,n)2$^4$He reaction rate 
reduces the CNO abundance\cite{Coc12a}, even though the \deu\ and \sep\ 
{\em final abundances are left unchanged}.
However, the greatest improvement comes from the Monte Carlo technique, now widely used in 
nucleosynthesis calculations \cite{Stats}. 
Ideally reaction rate uncertainties are known, together with the associated probability density functions (p.d.f.).
As described in Longland \zetal\ \cite{Eval1}, this can be obtained by Monte Carlo calculations taking into
account uncertainties and p.d.f. of experimentally (or theoretically) determined quantities that 
enter into the rate calculations. Reaction rate p.d.f. can usually be represented by a log-normal
distribution whose parameters are tabulated as a function of temperature \cite{Eval1}. 
These reaction rate p.d.f. can then be used in nucleosynthesis  Monte Carlo calculations
where all reaction rates are sampled independently. From the resulting histograms of 
calculated abundances, the median and 68\% confidence interval is obtained from the 0.5, 0.16 and 0.84
quantiles. This is how the confidence intervals quoted here are obtained.

Namely the reaction rates ${N_A}\langle\sigma{v}\rangle_k$, (with $k$ being the index of the reaction), 
are assumed to follow a lognormal distribution:
\begin{equation}
{N_A}\langle\sigma{v}\rangle_k=\exp\left(\mu_k(T)+p_k\sigma_k(T)\right)
\label{q:ln}
\end{equation}
where  $p_k$ is sampled according to a {\em normal} distribution of mean 0 and variance 1 (Eq.~(22)  of Ref.~\refcite{STARLIB}).
The $\mu_k$ and $\sigma_k$ determine the location of the distribution and its width, which are tabulated as a function of $T$.
First, by taking the  quantiles of the Monte Carlo calculated distributions of final isotopic abundances one obtains, not
only their median values but also the associated confidence interval.
Second, the (Pearson's) correlation coefficient between isotopic abundance $y_{j}$  and reaction rate random enhancement factors ($p_k$ in Eq.~\ref{q:ln}) can be calculated.
Figure~\ref{f:cor}, in the top panels, shows simple cases where there is no or a very strong 
correlation between a yield and two reaction rates. A simple sensitivity study would have been
sufficient here, but the bottom panels displays some weak correlation between the CNO production
and two reaction rates. These were not recognized in a study, changing each of these reaction rate, 
{\em one at a time}, by factors up to 1000 \cite{Coc12a}.
This explain that in a Monte Carlo BBN calculation of CNO abundance, in the resulting
distribution (Figure~4 of Ref.~\refcite{Coc14b}), for $\approx$2\% of the cases, CNO/H$>10^{-13}$, a value that 
may affect first stars throughout their evolution (Pop III stars). A {\em combination} of higher rates together with lower rates for a few reactions around $^{10}$Be (Fig.~\ref{f:netw})
lead to this effect \cite{Coc14b}. This was not seen without the use of Monte Carlo and correlation analyses;
a combined technique that can be extended to other sites of nucleosynthesis \cite{Par08}.

\section{Conclusion}

The agreement between BBN predictions and observations is quite satisfactory except for lithium.
Many studies have been devoted to the resolution of this lithium problem and many possible ``solutions'', none fully satisfactory, have been proposed. 
For a detailed analysis see \cite{Fie11} and the various contributions to the meeting ``Lithium in the cosmos''~\cite{Spi12}.
In particular nuclear physics solutions, leading to an increased $^7$Be  destruction, have been experimentally investigated,
and can now be excluded \cite{Ham13}. Now that the D/H primordial abundance is expected to be known
with an improved precision \cite{Coo14}, nuclear cross sections of all reactions leading to \deu\ destruction should be 
determined with an equal precision \cite{DiV14,Coc1X}. 

Nevertheless, primordial nucleosynthesis remains a invaluable tool for
probing the physics of the early universe. When we look back in time,
it is the ultimate process for which we {\it a priori} know all the 
physics involved: departure from its predictions provide hints
for new physics or astrophysics. 
Hence, there are two motivations to extend BBN beyond the standard model: use it to probe the early universe and to test 
fundamental physics \cite{Ioc09,Pos10} on the one hand, and
find a solution to the lithium problem \cite{Fie11} on the other hand. 

Gravity that could differ from its general relativistic description, affecting the rate of expansion of the universe
(see Ref.~\refcite{JPU10} for a review), or the variation of the fundamental constants (see Ref.~\refcite{JPU11} for a review), 
can be constrained by BBN \cite{Coc09,Coc12b}. 
The decay of a massive particle during or after BBN could affect the light element abundances and potentially lower the \sep\ abundance 
(see e.g. Ref.~\refcite{Cyb13} and references therein).
This effect could also be obtained with negatively charged relic particles, like the supersymmetric partner of the tau lepton, that could
form bound states with nuclei, lowering the Coulomb barrier and hence
leading to the catalysis of nuclear reactions (see e.g. Ref.~\refcite{Kus13b} and references therein). 
Non--standard solutions to the lithium problem include photon cooling \cite{Erk12b}, possibly combining 
particle decay and magnetic fields \cite{Yam14}.

Last but not least, we stress here the importance of sensitivity studies in nuclear astrophysics: even in the 
simpler context of BBN without the complexity (e.g. mixing) of stellar nucleosynthesis, it would have been
very unlikely to predict the influence of  the $^1$H(n,$\gamma)^2$H reaction on \sep\ nor
of  the $^7$Li(d,n)2$^4$He reaction on CNO, before these systematic investigations.

\section*{Acknowledgements}
We are indebted to our collaborators on these topics:  Pierre Descouvemont, Fa\"{\i}rouz Hammache,
St\'ephane Goriely, Christian Iliadis, Richard Longland,
Keith Olive, Patrick Petitjean,
Maxim Pospelov and Jean-Philippe Uzan.
This work made in the ILP LABEX (under reference ANR-10-LABX-63) was supported by French state funds managed by the ANR 
within the Investissements d'Avenir programme under reference ANR-11-IDEX-0004-02 and by the ANR VACOUL, ANR-10-BLAN-0510.


\begin{thebibliography}{0}

\bibitem{Wag73} R.V. Wagoner, 
\zapj\ {\bf 179} (1973) 343.

\bibitem{Yan79} J. Yang, D. Schramm, G. Steigman and R. T. Rood, 
\zapj\ {\bf 227}  (1979) 697.

\bibitem{LEP} 
The ALEPH Collaboration (S. Schael et al.) and The DELPHI Collaboration (J. Abdallah et al.) and the L3 Collaboration 
(M. Acciarri et al.) and The OPAL collaboration (G. Abbiendi et al.) and The SLD Collaboration (Kenji Abe et al.) and the 
LEP Electroweak Working Group and the SLD Electroweak and Heavy Flavour Groups, 
\zprep\ {\bf 427} (2006) 257. 

\bibitem{WMAP9} G. Hinshaw, D. Larson, E. Komatsu \etal,
\zapjs\ {\bf 208} (2013) 19.

\bibitem{Planck13} {\it{Planck  }}  Collaboration XVI, P.A.R. Ade, N. Aghanim, C. Armitage-Caplan, M. Arnaud, M. Ashdown \etal, 
\zaa\ {\bf 571} (2014) A16.

\bibitem{Planck15} {\it{Planck  }}  Collaboration XIII, P.A.R. Ade, N. Aghanim,  M. Arnaud, M. Ashdown, J. Aumont  \etal,
\zaa\  {\bf 594} (2016) A13.

\bibitem{Ioc09}F. Iocco, G. Mangano, G. Miele, O. Pisanti and P. D. Serpico,
\zpr\  {\bf 472}  (2009) 176.

\bibitem{Pos10} M. Polspelov and J. Pradler,
{\it Annual Review of Nuclear and Particle Science} {\bf 60} (2010) 539. 

\bibitem{Weinberg2008} S. Weinberg,
{\em Cosmology},
Oxford University Press, 2008, ISBN 978-0-19-852682-7.

\bibitem{Fuk04} M. Fukugita and P.J.E. Peebles,
\zapj\ {\bf 616} (2004) 643.

\bibitem{Coc14b} A. Coc,   J.--P. Uzan, and E. Vangioni
\zjcap\  {\bf 10} (2014) 050   {\tt arXiv:1403.6694 [astro-ph.CO]}.

\bibitem{Spi82} F.~Spite and M.~Spite,
\zaa, {\bf  115}  (1982) 357.

 \bibitem{Sbo10} L. Sbordone, P. Bonifacio, E. Caffau   \etal,
\zaa, {\bf 522} (2010) 26.

\bibitem{How12} J.C.  Howk,  N. Lehner, B.D. Fields and G.J. Mathews,  
\znat\ {\bf 489} (2012) 121.

\bibitem{Taj15} A. Tajitsu, K. Sadakane, H. Naito, A. Arai and W. Aoki,
\znat\ {\bf 518}  (2015) 381.

\bibitem{Izz15} L. Izzo, M. Della Valle, E. Mason, F. Matteucci, D. Romano \etal,
\zapj\ {\bf 808} (2015) L14.
 
\bibitem{HJCI96} M.~Hernanz, J.~Jos\'e, A.~Coc and J.~Isern,
\zapj\ {\bf 465} (1996) L27. 

\bibitem{Pet12} M. Pettini and M. Cooke, 
\zmnras\ {\bf 425} (2012) 2477.

\bibitem{Coo14} R. Cooke, M. Pettini, R.A. Jorgenson, M.T. Murphy and C.C. Steidel, 
\zapj\   {\bf 781} (2014) 31.

\bibitem{Coc1X} A. Coc, P. Petitjean, J.-P. Uzan, E. Vangioni,  P. Descouvemont, C. Iliadis and R.~Longland,
\zprd\ {\bf 92} (2015) 123526 {\tt arXiv:1511.03843 [astro-ph.CO]}. 

\bibitem{Ave15} E. Aver, K.A. Olive and  E.D. Skillman,
\zjcap\ 07 (2015) 011 {\tt arXiv:1503.08146v1 [astro-ph.CO]} 

\bibitem{Ban02}
T.~Bania, R.~Rood and D.~Balser, 
\znat\ {\bf 415} (2002) 54.

\bibitem{Dic82} 
D.~Dicus, E.~Kolb, A.~Gleeson, E.~Sudarshan, V.~Teplitz and M.~Turner, 
\zprd\ {\bf 26}  (1982) 2694.

\bibitem{PDG} 
K.A. Olive et al. (Particle Data Group), 
{\em Chin. Phys. C} {\bf 38} (2014) 090001 {\tt URL: http://pdg.lbl.gov}

\bibitem{Wie11} F. Wietfeldt and G. Greene, 
{\it Rev. Mod. Phys.} {\bf 83} (2011) 1173.

\bibitem{You14}
A.R. Young, S. Clayton, B.W. Filippone, P. Geltenbort, T.M. Ito, C.-Y. Liu, M. Makela, C.L. Morris, B. Plaster, A. Saunders, 
S.J. Seestrom and R.B. Vogelaar, {\it J. Phys. G: Nucl. Part. Phys.} {\bf 41}  (2014) 114007.

\bibitem{Mat05} G. J. Mathews, T. Kajino and T. Shima
\zprd\ {\bf 71} (2005) 021302(R). 

\bibitem{And06} S. Ando, R.H. Cyburt, S.W. Hong and C.H. Hyun,
\zprc\ {\bf 74} (2006) 025809.

\bibitem{Des04} 
P.~Descouvemont, A.~Adahchour, C.~Angulo, A.~Coc and E.~Vangioni--Flam, 
\zadndt\ {\bf 88}  (2004) 203.

\bibitem{Cyb04} R.H. Cyburt, 
 \zprd\ {\bf 70}  (2004)  023505.

\bibitem{Ser04} P.D.~Serpico, S.~Esposito, F.~Iocco, G.~Mangano, G.~Miele and O.~Pisanti,
\zjcap\  {\bf 12}  (2004) 010.

\bibitem{Coc12a} A. Coc., S. Goriely, Y., Xu, M. Saimpert and E. Vangioni, 
 \zapj\ {\bf 744}  (2012) 158.

\bibitem{DiL09} A. Di Leva, L. Gialanella, R. Kunz \etal,
\zprl\ {\bf 102}  (2009) 232502. 

\bibitem{Cyb08a} R.H. Cyburt and B.~Davids,
 \zprc\ {\bf 78}  (2008) 064614.
 
 \bibitem{deB14} R. J. deBoer, J. G\"orres, K. Smith, E. Uberseder, M. Wiescher, 
 A. Kontos, G. Imbriani, A. Di Leva and F. Strieder,
\zprc\ {\bf 90} (2014) 035804.

\bibitem{DiV14} E. Di Valentino, C. Gustavino, J. Lesgourgues, G. Mangano, A. Melchiorri, G. Miele and O. Pisanti,
\zprd\ {\bf 90} (2014) 023543.

 \bibitem{CV10} A.~Coc and E.~Vangioni,  
{\em Journal of Physics Conference Series}, {\bf 202}  (2010) 012001.

\bibitem{Leo06} 
D.S. Leonard, H.J. Karwowski, C.R. Brune, B.M. Fisher  and E.J. Ludwig,
\zprc\ {\bf 73} (2006) 045801.

\bibitem{Ara11} K. Arai, S. Aoyama, Y. Suzuki, P. Descouvemont and D. Baye,
\zprl\  {\bf 107}  (2011) 132502.

\bibitem{Mar05} L.E. Marcucci, M. Viviani, R. Schiavilla, A. Kievsky and S. Rosati,
\zprc\ {\bf 72}  (2005) 014001; Laura Elisa Marcucci {\it priv. comm.}. 

\bibitem{Hou15} S.Q. Hou, J.J. He, S. Kubono and Y.S. Chen,
\zprc\ {\bf 91} (2015) 055802 {\tt arXiv:1502.03961 [astro-ph.CO]}.

\bibitem{Cyb08b} R.H. Cyburt, B.D. Fields and K.A. Olive,
\zjcap\ {\bf 11}  (2008)  012.

\bibitem{Cyb15} R.H. Cyburt, B.D. Fields, K.A. Olive and T.-H. Yeh,
\zrmp\ {\bf 88}  (2016) 015004 {\tt arXiv:1505.01076 [astro-ph.CO]}.

\bibitem{Ham10} F.~Hammache, M.~Heil, S.~Typel et al.
\zprc\  {\bf 82} (2010) 065803 .

\bibitem{Asp06} M.~Asplund, D.~Lambert, P.E. F.~Nissen, F.~Primas and V.~Smith, 
\zapj, {\bf 644}  (2006) 229.

\bibitem{Lin13} K. Lind, J. Melendez, M. Asplund, R. Collet and Z. Magic,
\zaa\ {\bf 554}  (2013) A96.

\bibitem{Fie11}  B. Fields,  
{\em Annu. Rev. Nucl. Part. Sci.}  {\bf 61}  (2011) 47.

\bibitem{Spi12} M. Spite, F. Spite and P. Bonifacio
{\em Mem. S.A.It. Suppl.} {\bf 22}  (2012) 9.

\bibitem{Fu15} X. Fu, A. Bressan, P. Molaro and P. Marigo,
\zmnras\ {\bf 452}, (2015) 3256
{\tt arXiv:1506.05993v1 [astro-ph.SR]}.

\bibitem{Coc04} A.~Coc, A., E.~Vangioni-Flam, P.~Descouvemont, A.~Adahchour and C.~Angulo, 
\zapj\ {\bf 600}  (2004) 544.

\bibitem{Kir11} O.S.~Kirsebom and B.~Davids, 
\zprc\ {\bf 84} (2011) 058801.

\bibitem{Cha11} N.~Chakraborty,  B.D.~Fields and K.A.~Olive, 
\zprd\ {\bf 83}   (2011) 063006.

\bibitem{Ham13} 
F.~Hammache, A.~Coc, N.~de S\'er\'eville \etal, 
\zprc\  {\bf 88}  (2013) 062802(R) {\tt arXiv:1312.0894 [nucl-ex]}.

\bibitem{Ren88}M. H. Reno and D. Seckel,
\zprd\ {\bf 37}  (1988) 3441.

\bibitem{Jed04} K.~Jedamzik,
\zprd\ {\bf 70} (2004) 063524.

 \bibitem{Ber01} Z. Berezhiani,  D. Comelli and F.L. Villante,  
 \zplb\ {\bf 503} (2001) 362.

\bibitem{Coc14a} A.~Coc,  M. Pospelov, J.-Ph.~Uzan and E.~Vangioni
\zprd\ {\bf 90} (2014)  085018  {\tt arXiv:1405.1718 [astro-ph]}.

\bibitem{Kus14} M. Kusakabe, M.-K. Cheoun and K. S. Kim,
\zprd\ {\bf 90} (2014) 045009.

\bibitem{Oli12} 
K.~Olive, P.~Petitjean, E. Vangioni and J. Silk, 
\zmnras, {\bf 426}  (2012) 1427.

\bibitem{Dai06} F. Daigne, K.A. Olive, J. Silk, F. Stoehr and E. Vangioni,  
\zapj\ {\bf 647}  (2006) 773. 

\bibitem{Rol09} E. Rollinde, E. Vangioni, D. Maurin, K.A. Olive, F. Daigne, J. Silk, 
and  F.H. Vincent, 
\zmnras\ {\bf 398}  (2009) 1782. 

\bibitem{Van14} E. Vangioni, K.A. Olive, T. Prestegard, J. Silk, P. Petitjean and V. Mandic, 
\zmnras\ {\bf 447}  (2014)  2575 {\tt arXiv:1409.2462 [astro-ph.GA]}.

\bibitem{press74} W.H. Press \& P. Schechter Astrophys. J. {\bf 187} (1974) 425.

\bibitem{Lin06} J.L. Linsky, B.T. Draine, H.W. Moos, \etal,
\zapj\ {\bf 647}  (2006) 1106.

\bibitem{Pro10} T. Prodanovi\'c, G. Steigman and B. D. Fields,
\zmnras\. {\bf 406}  (2010) 1108. 

\bibitem{oesch14} P.A. Oesch, R.J. Bouwens, and G.D. Illingworth, et al., Astrophys. J. {\bf 786} (2014) 108.

\bibitem{bou14} R. J. Bouwens, L. Bradley, A. Zitrin, et al., Astrophys. J. {\bf 795} (2014) 126.

\bibitem{bou15} R. J. Bouwens, R.J. Illingworth, G.D. Oesch, et al., Astrophys. J. {\bf 803} (2015) 34.

\bibitem{Sal55} E.E. Salpeter, 
\zapj\ {\bf 121}, 161 (1955).

\bibitem{Stats} C. Iliadis, R. Longland, A. Coc, F.X. Timmes \& A. E. Champagne,
\zjpg\ : {\it Nucl. Part. Phys.}, {\bf 42} (2015) 034007 {\tt arXiv:1409.5541 [nucl-ex]}.

\bibitem{Eval1} R. Longland, C. Iliadis, A.~E. Champagne \etal, 
\znpa\  {\bf 841} (2010) 1. 

\bibitem{STARLIB}  A.L. Sallaska, C. Iliadis, A.E. Champagne S. Goriely, S. Starrfield,  F. X. Timmes, 
\zapjs\ \textbf{207}  (2013)  18 {\tt arXiv:1304.7811 [astro-ph.SR],
http://starlib.physics.unc.edu/index.html}.

\bibitem{Par08} A. Parikh, J. Jos\'e, F. Moreno, C. Iliadis,
\zapjs\ \textbf{178}  (2008) 110. 

\bibitem{JPU10}  J.-P. Uzan, 
{\em General Relativity and Gravitation},  {\bf 42} (2010) 2219.

\bibitem{JPU11} J.-P. Uzan, 
{\em Living Reviews in Relativity}, {\bf 14}  (2011) 2. 

\bibitem{Coc09} A. Coc, K. Olive, J.-P. Uzan and E. Vangioni
\zprd\ {\bf 79}   (2009) 103512. 

\bibitem{Coc12b}  A. Coc, P. Descouvemont, K. Olive, J.--P. Uzan and E. Vangioni, 
\zprd\ {\bf 86}  (2012) 043529. 

\bibitem{Cyb13}  R. H. Cyburt, J. Ellis, B. D. Fields, F. Luo, K. A. Olive and V. C. Spanos
\zjcap\ {\bf 05}  (2013) 014. 

\bibitem{Kus13b} M. Kusakabe, K. S. Kim, M.-K. Cheoun, T. Kajino and Y. Kino,
\zprd\ {\bf 88}  (2013) 063514.

\bibitem{Erk12b} O. Erken, P. Sikivie, H. Tam and Q. Yang
\zprd\  {\bf 85}  (2012) 063520. 

\bibitem{Yam14} D. G. Yamazaki, M. Kusakabe, T. Kajino, G.J. Mathews and M.K. Cheoun,
\zprd\ {\bf 90} (2014) 023001.


\end{thebibliography}
\end{document}